\documentstyle[11pt,epic,eepic]{article}
\newcommand{\ol}{\setlength{\itemsep}{0pt.}\begin{enumerate}}
\newcommand{\eol}{\end{enumerate}\setlength{\itemsep}{-\parsep}}

\newtheorem{theorem}{Theorem}[section]

\newtheorem{lemma}[theorem]{Lemma}

\newcommand{\qedsymb}{\hfill{\rule{2mm}{2mm}}}
\newenvironment{proof}{\begin{trivlist}
\item[\hspace{\labelsep}{\bf\noindent Proof: }]
}{\qedsymb\end{trivlist}}

\newcommand{\remove}[1]{}

\newcounter{alpha:count}
{\begin{list}{{\alph{alpha:count}.}}%
{\usecounter{alpha:count}\setlength{\listparindent}{0pt}}}%
{\end{list}}

\newcount\hour  \newcount\minutes  \hour=\time  \divide\hour by 60
\minutes=\hour  \multiply\minutes by -60  \advance\minutes by \time
\def\Month{\ifcase\month\or January\or February\or March\or April\or May\or 
June\or July\or August\or September\or October\or November\or December\fi}

\def\Minutes{\ifnum\minutes<10 0\fi\number\minutes}

\begin{document}


\renewcommand{\thefootnote}{\fnsymbol{footnote}}

\title{Dijkstra's Self-Stabilizing Algorithm in Unsupportive Environments\\
{\large (Preliminary Version)}}

\author{Shlomi Dolev\\
Dept. of Computer Science\\
Ben-Gurion University --- Israel\\
\and
Ted Herman\\
Dept. of Computer Science\\
University of Iowa --- USA}

\date{19 April 2001}  

\maketitle

\renewcommand{\thefootnote}{\arabic{footnote}}

\begin{abstract}
The first self-stabilizing algorithm \cite{Dij73}
assumed the existence of a central daemon,
that activates one processor at time
to change state as a function of its own state
and the state of a neighbor.  Subsequent research
has reconsidered this algorithm without the
assumption of a central daemon, and under different
forms of communication, such as the model of link
registers.  In all of these investigations, one 
common feature is the atomicity of communication, 
whether by shared variables or read/write registers.
This paper weakens the atomicity assumptions for
the communication model, proposing versions of \cite{Dij73}
that tolerate various weaker forms of atomicity.
First, a solution for the case of regular registers
is presented.  Then the case of safe registers is considered,
with both negative and positive results presented.
The paper also presents an implementation of \cite{Dij73}
based on registers that have probabilistically correct
behavior, which requires a notion of weak stabilization.
\end{abstract}

\section{Introduction}
The self-stabilization concept is not tied to 
particular system settings. 
Our work considers several new system settings 
and demonstrates the applicability of the 
self-stabilization paradigm to these systems.
In particular, we investigate systems with
regular and safe registers and present modifications
of Dijkstra's first self-stabilizing algorithm \cite{Dij73} that
stabilizes in these systems.

The solution for the regular registers case use a special 
label in between writes of labels. In the case of safe registers
we prove impossibility results, for the cases in
which neighboring processors use a single safe register 
to communicate between themselves --- where the register 
is/isn't divided to multiple fields. 
In the positive side, we define a composite safe
register that roughly speaking ensures that reads return
at most one corrupted field and design an algorithm for that
case. Then we allow the processors to read the value written in their
registers (therefore avoiding extra writes for refreshes).
We present two algorithms for the above case, one that uses 
unary encoding and another that is based on Gray code.

Then we introduce {\it randomized registers} that, roughly speaking, return 
the ``correct value'' with probability $p$. 
It is impossible to ensure closure in such a system,
since all reads may return incorrect values.
We introduce the notion of {\it weak self-stabilization} for
such systems. We use Markov chains to compute the ratio
between the number of safe configurations and unsafe
configurations in an infinite execution.

Markov chains associate each state (system configuration)
with a probability to be in this state during an infinite 
execution. The fixed probability of the state is a ``stabilizing''
value. It is clear that the probability is either zero or one
in the first configuration. Given the probability of transitions
between configurations, one can compute the stable probability
in an infinite execution, which is typically greater than zero
and less than one. We found the definition of weak stabilization
and the use of Markov chains to be an interesting and
promising way for extending the applicability of the
self-stabilizing concept.

The remainder of the paper is organized as follows.
In the next section we describe a solution for regular registers.
Then in Section \ref{s:sr} we present impossibility results 
and algorithms for different settings of systems that use
safe registers. Randomized registers and the use of Markov chains
are presented in Section \ref{s:ra}. Detailed proofs  are omitted
from this extended abstract.

\section{Regular Registers}
\label{s:rr}

Before we introduce our results for the case of regular registers
let us presents ``folklore'' results concerning read/write registers.

\noindent
{\bf Read/Write Atomicity:} It is known that $n-1$ labels are 
sufficient for the convergence of Dijkstra algorithm assuming
a central daemon, where $n$ is the number of processors in the ring.
We next prove that $n-2$ labels are not sufficient.

\noindent
{\bf Lower bound:} Consider the case of $n-2$ states in a system of 
$n=5$ processors. Thus there are three possible processor states,
which we label \{0,1,2\}.  To prove impossibility we demonstrate
a non-converging sequence of transitions (the key to constructing
the sequence is to maintain all three types of labels in each 
system state, which violates the key assumption for the proof
of convergence).

$\{0,0,2,1,0\}$ 
$\stackrel{p_1}{\rightarrow}$ 
$\{1,0,2,1,0\}$
$\stackrel{p_5}{\rightarrow}$ 
$\{1,0,2,1,1\}$ 
$\stackrel{p_4}{\rightarrow}$ 
$\{1,0,2,2,1\}$ 
$\stackrel{p_3}{\rightarrow}$ 
$\{1,0,0,2,1\}$ 
$\stackrel{p_2}{\rightarrow}$ 
$\{1,1,0,2,1\}$.

We now present a reduction (see \cite{Do2000})
of a ring with $2n$ processors that is activated 
by a central daemon to a ring with $n$ processors that
assumes read write atomicity. We conclude that at least
$2n-1$ states are required.

Each processor 
$p_j$ has an internal variable
in which $p_j$ stores the value $p_j$ reads from $p_{j-1}$.
Each read is a copy to an internal variable and each write 
is a copy of internal variable to a register.
Thus, we have in fact a ring of $2n$ processors 
in a system with a central daemon. Hence, 
$2n-1$ states are required and are sufficient.

We now turn to design an algorithm
for the case of regular registers.
Informally, a regular register has the property
that a read operation concurrent with a write
operation can return either the ``old'' or ``new''
value.  More formally, 
to define a regular register $r$ we need
to define the possible values that a read operation
from $r$ returns. 
Let $x^0$ be the value of the last write operation
in $r$ that ends prior to the beginning of the read operation
(let $x^0$ be the initial value of $r$ if no such write exists).

A read operation from a regular register $r$ that
is not executed concurrently with a write operation
to $r$ returns $x^0$.
A read operation from a regular register $r$ that
is executed concurrently with a write of a value $x^1$
returns either $x^0$ or $x^1$.  Note that more generally, 
a read concurrent with a sequence of write operations 
of the values $x^1,x^2,\cdots$ to $r$ could return 
any $x^k$, however once a read returns $x^k$ for $k>1$, no subsequent
read by the same reader will return $x^j$ for $j<k-1$.   

A naive implementation of Dijkstra's algorithm using regular registers
may result in the following execution:

\begin{figure} [ht]
\begin{center}
\fbox{
\setlength{\textwidth}{6.00in}
\begin{minipage}{\textwidth}         
\begin{tabbing}
xx \= xx \= xx\= xxxxxxx \= xxxxxxx \= xxxxxxx \= xxx \= xxx \= xxx \= \kill
$s_1$ \> $s_2$ \> $s_3$  \\
0 \> 0 \> 0 \> $p_1$ starts to write 1\\
1 \> 0 \> 0 \> $p_1$ still writing 1, and $p_2$ reads 1\\ 
1 \> 1 \> 0 \> $p_1$ still writing 1, and $p_2$ writes 1\\
1 \> 1 \> 0 \> $p_1$ still writing 1, $p_2$ writes 1, and $p_3$ reads 1 \\
1 \> 1 \> 0 \> $p_1$ still writing 1, $p_2$ reads 0 \\
1 \> 0 \> 1 \> $p_2$ writes 0, and $p_3$ writes 1 \\
1 \> 0 \> 1 \> $p_1$ reads 1, $p_2$ reads 1, and $p_3$ reads 0
\end{tabbing}
\end{minipage}
}
\caption[]{
\label{f:regularI}
Straightforward regular register implementation fails}
\end{center}
\end{figure}

We have started in a safe configuration in which all 
the values (in the registers and the internal variables)
are $0$ and we have reached a configuration in which all the
processors may simultaneously change a state.

To overcome the above difficulty we introduce a new value
$\perp$ that is written before any change of a value of a
register. The algorithm for the case of regular registers
appears in Figure \ref{f:DKr}.  In the figure, ${\sf IR}_i$
is the input register for $p_i$ (thus ${\sf IR}_i$
is the output register of $p_{i-1}$).  Variable $x_i$
contains the counter defined for Dijkstra's algorithm, and
variable $t_i$ is introduced to emphasize the fine-grained
atomicity of the model (one step reads a register,
and the value it returns is tested in another step). 

A safe configuration is a configuration in which all the 
registers have the same value, say $x$, and every read operation
that has already started will return $x$.
For simplicity we assume there are $2n+1$ states.
Therefore,
it is clear that a state is missing in the initial 
configuration,
say the state $y$. Hence, when
$p_1$ writes $y$,  $p_1$ does not change its state
before reading it from $p_n$. $p_n$ can read $y$
only when $p_{n-1}$ has the state $y$.
Any read operation of $p_{n-1}$ that starts following the 
write operation that assigns $y$ to $p_{n-1}$
may return either $\perp$ or $y$, which is 
effectively $y$ (see lines 3 to 6 and 10 to 13 
of the code).

\begin{figure} [ht]
\begin{center}
\fbox{
\setlength{\textwidth}{6.00in}
\begin{minipage}{\textwidth}         

\begin{tabbing}
xx \= xxxxxxxx \= x \= xxx \= xxx \= xxx \= xxx \= \kill
1 \> $p_1$: \> \>{\bf do} forever \\
2 \> \> \> \>{\bf read} $t_1 := {\sf IR}_1$ \\
3 \> \> \> \>{\bf if} $t_1=x_1$ {\bf then} \\
4 \> \> \> \> \> \>$x_1~:=~(x_1~+~1)$ mod $(2n~+~1)$\\
5 \> \> \> \> \> \>{\bf write} ${\sf IR}_2~:=~\perp$\\
6 \> \> \> \> \> \>{\bf write} ${\sf IR}_2~:=~x_1$\\
7 \> \> \> \>{\bf else} {\bf write} ${\sf IR}_2~:=~x_1$ \\ 
8 \> $p_i$ ($i \neq 1$): \> \> {\bf do} forever\\
9 \> \> \> \>{\bf read} $t_i := {\sf IR}_i$ \\
10 \> \> \> \>{\bf if} $x_i \neq t_i$ {\bf then}\\
11 \> \> \> \> \> $x_i~:=~t_i$\\
12 \> \> \> \> \>{\bf write} ${\sf IR}_{i+1}~:=~\perp$\\
13 \> \> \> \> \>{\bf write} ${\sf IR}_{i+1}~:=~x_i$\\
14 \> \> \> \>{\bf else} {\bf write} ${\sf IR}_{i+1}~:=~x_i$ 
\end{tabbing}

\end{minipage}
}
\caption[]{
\label{f:DKr}}
Dijkstra's Algorithm for Regular Registers
\end{center}
\end{figure}

\section{Safe Registers}
\label{s:sr}

Safe registers have the weakest properties of any in Lamport's
hierarchy.  A read concurrent with a write to a safe register 
can return any value in the register's domain, even if the value
being written is already equal to what the register contains.
There are two cases to consider for the model of safe registers.
If a processor is unable to read the register(s) that it writes,
we can show that Dijkstra's algorithm cannot be implemented.  We
initially consider the model of a single link register for each
processor under the restriction that a writer is unable to read
its output registers.  

\begin{lemma} 
Dijkstra's algorithm cannot be implemented using only 
a single 1W1R safe register between $p_i$ and $p_{i+1}$.
\end{lemma}
\begin{proof}
Processor $p_i$ ($i \neq 1$) that copies from the
output register of $p_{i-1}$ must continually rewrite
its output register for $p_{i+1}$ --- otherwise
there can be a deadlock where the value written
by $p_i$ is different from the value $p_i$ reads
from $p_{i-1}$. Similarly, $p_1$ must repeatedly write, otherwise 
there can be a deadlock where all the registers
have the same value, and the $p_1$ program counter is past 
the first write to its register that incremented this value.
Therefore, processors continually write into their output registers.
Since all processors repeatedly write their output registers,
we can construct an execution where reads are concurrent
with writes and obtain arbitrary values.  This construction
can be used to show that the protocol does not converge (and
also that it is not stable).
\end{proof}
  
\noindent
{\bf Multiple fields safe register:}
The next result we present is impossibility for the case of multiple 
safe registers per processor, but where processors cannot read the registers
they write.  Suppose each processor $p_i$ has $m$ safe registers
to write, which $p_{i+1}$ reads, and also $p_i$ reads $m$ safe registers
written by $p_{i-1}$.  If a protocol allows
a state in which a processor does not write any of its registers so long
as its state does not change, then we may construct a deadlock because
the local state of the processor differs from the encoding of values 
contained in its output registers.  Therefore, in any implementation of
the protocol, we can construct an execution fragment so that any chosen 
processor $p_i$ writes at least some of its registers $t$ times, for
arbitrary $t>0$, and during the same execution fragment, 
$p_{i-1}$ takes no steps.  
Moreover, if $p_i$ does not write to all $m$ registers, then the registers
it does not write can have arbitrary values inherited from the initial 
state.  Therefore, $p_{i+1}$ can read any value from $p_i$, 
since at each step of $p_{i+1}$
reading one of the $m$ registers written by $p_i$, we can construct
an execution in which $p_i$ is concurrently writing to the same safe 
register.  Because $p_{i+1}$ can read any value, it is possible that
for $i\neq n$ that $p_{i+1}$ reads a value equal to its own current value,
which for Dijkstra's algorithm, means that $p_{i+1}$ will maintain its
current value rather than changing it; for the case $i=n$, there is 
an execution where each time $p_1$ reads its input registers, the 
value read differs from its own value, and again $p_1$ makes no change
to its current value.  These situations can repeat indefinitely with 
no processor entering the critical section.

\noindent
{\bf Composite safe register:}
Next we sketch a solution in which fields of the registers
can be written and the entire register is read
{\it at once}. We call such a register {\it composite
safe register}.
A read from a composite safe register may return 
an arbitrary value for at most one of the register
fields, a field in which a write is executed 
concurrently to the read\footnote{This assumption reflects 
reality in system in which
a read operation is much faster than a write
operation.}. We note that there is a natural extension of our
algorithm in which at most
$k$ fields of a register may return an arbitrary value.

Each bit of the label value is stored in three 1-bit safe 
registers (three fields). This will ensure that a read during
a refresh operation will return the value of the register.
Assume that the value 101 is stored in nine 1-bit
safe registers as 111000111. Assume further that
a processor refreshes the value written in these
registers each time writing in one of the 1-bit safe
registers. A read operation returns the value of the
entire composite safe register in which at most one
bit is wrong. The Hamming distance ensures that
the original value of the label bit can be determined.

To allow a value change we add a three bits 
guard value. Hence, the composite safe register 
has three bits that function as a guard value and
$3\times 2 (n+1)$ bits for the label.

A processor $p_i$, $i\neq 1$, that reads a new value
from $p_{i-1}$ first sets the guard value to 0
(writing 000 in the guard bits),
and then changes the value of the label.
$p_i$ writes 111 to its guard bits once 
$p_i$ finishes updating the label.

A processor $p_i$ that reads a guard value $0$ does not
use the value read. When $p_i$ reads a guard
value $1$ it examines the value it read.

The correctness proof starts in convincing ourselves
that after the first time a processor $p_i$ refreshes
(or writes a new value in) its register any read operation
from its register (that returns a value) 
results in the last value written
to this register.
$p_1$ eventually writes a non existing label, this label
cleans the system. More details are omitted from this extended
abstract.

\noindent
{\bf Safe registers with reads instead of refreshes:} 
Given the above impossibility results, 
we examine settings where a processor can read the
contents of the registers in which it writes.
Consider $2n+1$ single bit, safe, 1W2R registers rather than a 
single register per processor.  Each processor maintains a 
counter with domain $[1,2n+1]$ for Dijkstra's algorithm.
Unary encoding represents this counter:  for a counter value 
$k$, the proper encoding is to write all registers 0 except
for the register with index $k$, which has value 1.  

\begin{figure} [ht]
\begin{center}
\fbox{
\setlength{\textwidth}{6.00in}
\begin{minipage}{\textwidth}         

\begin{tabbing}
xx \= xxxxxxxx \= x \= xxx \= xxx \= xxx \= xxx \= \kill
1 \> $p_1$: \> \>{\bf do} forever \\
2 \> \> \> \> {\bf do} $k:=1$ {\bf to} $2n+1$ \\
3 \> \> \> \> \> {\bf if} $k\neq x_1 \quad\wedge\quad {\sf IR}_2[k]=1$ \\
4 \> \> \> \> \> \> {\bf write} ${\sf IR}_2[k] := 0$ \\
5 \> \> \> \> \> {\bf if} $k = x_1 \quad\wedge\quad {\sf IR}_2[k]=0$ \\
6 \> \> \> \> \> \> {\bf write} ${\sf IR}_2[k] := 1$ \\
7 \> \> \> \> {\bf do} $s:= 0, \; j:=0, \; k:=1$ {\bf to} $2n+1$ \\
8 \> \> \> \> \> {\bf if} ${\sf IR}_1[k]=1$ \\
9 \> \> \> \> \> \> $s := s + 1; \; j:= k$ \\
10 \> \> \> \> {\bf if} $s=1 \quad\wedge\quad j=x_1$ \\
11 \> \> \> \> \> $x_1 := 1 + x_1 \bmod (2n+1)$ \\ \\
1 \> $p_i$ $(i\neq 1)$: \> \>{\bf do} forever \\
2-8 \> \> \> \>  ({\em same as for $p_1$}) \\ 
10 \> \> \> \> {\bf if} $s=1 \quad\wedge\quad j\neq x_i$ \\
11 \> \> \> \> \> $x_i := j$ \\
\end{tabbing}

\end{minipage}
}
\caption[]{
\label{f:DKc}}
Dijkstra's Algorithm for Safe Registers 
\end{center}
\end{figure}

A legitimate configuration for this protocol is that each register
vector represents the processor's last counter value (it differs
only when a processor updates its counter) and counters correspond
to Dijkstra's algorithm.  

\begin{lemma} \label{saf-lem} Figure \ref{f:DKc} is a self-stabilizing
implementation of Dijkstra's algorithm.
\end{lemma}
\begin{proof}  There are two proof obligations, stability (closure)
from legitimate configurations and convergence from arbitrary configurations
to legitimate ones.
 
\noindent \textbf{Closure.} 
It is straightforward to verify that in 
any processor cycle from a legitimate configuration, 
a processor writes to at most two registers as it changes the 
counter value.  Thus when the neighbor reads these registers, at most
two reads can have incorrect values due to concurrent writing.  If
both have correct values, the token passes correctly (a subsequent read
by the process can still obtain an incorrect value, but only by 
getting 0 for all reads, which causes no harm).  If both have incorrect
values, then the reader observes no change in counter values.  If just
one returns an incorrect value, then the reader observes parity of zero,
which is harmless.  This reasoning shows that the protocol is stable.

\noindent
\textbf{Convergence.} The remaining task is to verify that the
protocol guarantees to reach a legitimate configuration in any 
execution.  Suppose all processors have completed at least one
cycle of statements 1-11.  In the subsequent execution, a processor
only writes a register if that register requires change to agree with 
the processor's counter.  Note that by standard arguments, no deadlock
is possible in this system and that $p_1$ increments its counter 
infinitely many times in an execution.  
It is still possible that one processor 
can read more than two incorrect values due to concurrent writes (consider
an initial state with many counter values;  as these values are propagated
to some $p_i$, it could be that $p_{i+1}$ happens to read many registers
concurrent with $p_i$ writing to them).  Since the counter range is 
$[1,2n+1]$ and there are $n$ processors, it follows that at least one
counter value $t$ is not present in the system.  By the arguments given
for the proof of closure, no processor incorrectly reads input registers
to get the value $t$ in such a configuration.  Because $p_1$ increments
$x_1$ infinitely, we can suppose $x_1=t$ but no other processor or 
register encoding equals $t$, and by standard arguments (and the 
propagation of values observed in the proof of closure), a legitimate
configuration eventually is reached. 
\end{proof}

The protocol of Figure \ref{f:DKc} uses an expensive encoding
of counter values, requiring $2n+1$ separate registers.  The argument
for closure shows that changing a counter and transmitting it is 
effectively an atomic transfer of the value --- once the new value
is observed, then any subsequent read of the registers either 
returns the new value or some invalid value (where the sum of 
bits does not equal 1), which is ignored.  Note that this technique
\emph{is not} a general implementation of an atomic register from 
safe registers;  it is specific to the implementation of Dijkstra's
algorithm.  

Can we do better than using $2n+1$ registers?  The following 
protocol uses the Gray code representation of the counter, plus
a extra bit for parity.  The number of registers per processor is $m+1$ 
where $m=\lceil \lg (2n+1)\rceil$.
 
\begin{figure} [ht]
\begin{center}
\fbox{
\setlength{\textwidth}{6.00in}
\begin{minipage}{\textwidth}         

\begin{tabbing}
xx \= xxxxxxxx \= x \= xxx \= xxx \= xxx \= xxx \= \kill
1 \> $p_1$: \> \>{\bf do} forever \\
2 \> \> \> \> {\bf do} $k:=1$ {\bf to} $m$ \\
3 \> \> \> \> \> {\bf if} ${\sf IR}_2[k]\neq {\it Graycode}(x_1)[k]$ \\
4 \> \> \> \> \> \> {\bf write} ${\sf IR}_2[k] := {\it Graycode}(x_1)[k]$ \\
5 \> \> \> \> {\bf if} ${\sf IR}_2[m+1]\neq{\it parity}({\it Graycode}(x_1))$\\
6 \> \> \> \> \> {\bf write} ${\sf IR}_2[m+1] :={\it parity}({\it Graycode}(x_1))$ \\ 
7 \> \> \> \> {\bf do} $k:=1$ {\bf to} $m$ \\
8 \> \> \> \> \> $g[k] := {\sf IR}_1[k]$ \\
9 \> \> \> \> {\bf if} ${\sf IR}_1[m+1]={\it parity}(g) \quad\wedge\quad 
	Graycode(x_1)=g$ \\
10 \> \> \> \> \> $x_1 := (x_1 + 1) \bmod (2n+1)$ \\ \\
1 \> $p_i$ $(i\neq 1)$: \> \>{\bf do} forever \\
2-8 \> \> \> \>  ({\em same as for $p_1$}) \\ 
10 \> \> \> \> {\bf if} ${\sf IR}_i[m+1]={\it parity}(g) \quad\wedge\quad 
	Graycode(x_i)\neq g$ \\
11 \> \> \> \> \> $x_i := g$ \\
\end{tabbing}

\end{minipage}
}
\caption[]{
\label{f:DKg}}
Dijkstra's Algorithm using Gray Code 
\end{center}
\end{figure}

\begin{lemma} Figure \ref{f:DKg} is a self-stabilizing
implementation of Dijkstra's algorithm.
\end{lemma}
\begin{proof}
The closure argument is the same as given in the 
proof of Lemma \ref{saf-lem}, inspecting each of the four
cases of reading overlapping with writing of the two bits
that change when a processor changes its counter and
writes the one new Gray code bit and the parity bit.  In
each case, the neighbor processor either reads the old 
value, or ignores the values it reads (because parity is incorrect),
or obtains the new counter value.   The change from old to 
new counter value is essentially atomic.

Proof of convergence requires new arguments.  Consider some configuration
of an execution prior to which each processor has completed at least
two cycles of statements 1-11 in Figure \ref{f:DKg}, so that output
registers agree with counter values (unless the processor has read a
new value and updated its counter).  Observe that thereafter,
if processor $p_i$ successively reads two different Gray code values 
from its input registers, each with correct parity, then $p_{i-1}$ 
concurrently wrote at least once to its output registers.  Moreover, 
if $p_i$ successively reads $k$ different Gray code values with 
correct parity, then $p_{i+1}$ wrote at least $k$ times a new counter
value and read at least $k-1$ times from its own input registers, 
by the structure of the loop (statements 1-11) in Figure \ref{f:DKg}.
A consequence of these observations is that if $p_1$ successively reads
$k$ different counter values with correct parity, then $p_{n-k}$ wrote
at least one new counter value in the same period.  In particular, if
$p_1$ successively reads $n+2$ different counter values, then we may
assert that $p_2$ read $p_1$'s output registers and wrote a new counter
in the same period.  By the standard argument refuting deadlock, processor
$p_1$ increments its counter infinitely often in any execution.  Therefore
we can consider an execution suffix starting with $x_1=0$.  In the typical 
reflected Gray code, the high-order bit starting from $x_1=0$ does not
change until the counter has incremented $2^m$ times.  Therefore, 
until $p_1$ has incremented $x_1$ at least $2^m$ times, any read by
$p_2$ obtains a value with zero in the high-order bit.  The observations
above imply that, before $x_1$ changes at the high-order bit, each
processor has copied some counter value obtained via $p_1$ --- such 
counter values may be inaccurate due to reads overlapping writes or 
more than one write (bit change) for one scan of a set of registers,
however the value for the high-order bit stabilizes to zero in this
execution fragment.  In a configuration where no counter or register 
set has 1 in the high-order bit, the event of $p_1$ changing the 
high-order bit creates a unique occurrence of 1 in that position.
Since $p_1$ does not again change its counter until observing the 
same value from $p_n$, convergence is guaranteed.
\end{proof}

\section{Randomized State Reads and Weak Stabilization}
\label{s:ra}
Consider a system with a {\it fair central daemon}, 
in any given configuration
the daemon activates each of the processors with equal probability.
A system is {\it weakly stabilizing} if, in any execution, 
the probability that the system remains in any set of illegitimate
configurations is zero.  This definition implies that a weakly 
stabilizing system has the property that its state is infinitely
often legitimate. In addition, one can sum up the probabilities for being in a
legitimate state and use this value to compare 
algorithms.

To apply the definition of weak stabilization, we model 
register behavior probabilistically:
a processor that makes a transition may ``read'' an incorrect 
value and therefore make an errant transition.
We use Markov chains to analyze the percentage of the execution
in which the system will not be in a safe configuration.
See \cite{Lu79} for a description of Markov chains.

\begin{figure}
    \makeatletter     
\begin{center}
\setlength{\unitlength}{0.00083333in}
\begingroup\makeatletter\ifx\SetFigFont\undefined%
\gdef\SetFigFont#1#2#3#4#5{%
  \reset@font\fontsize{#1}{#2pt}%
  \fontfamily{#3}\fontseries{#4}\fontshape{#5}%
  \selectfont}%
\fi\endgroup%
{\renewcommand{\dashlinestretch}{30}
\begin{picture}(5992,5809)(0,-10)
\put(1800,3804){\makebox(0,0)[lb]{\smash{{{\SetFigFont{12}{14.4}{\rmdefault}{\mddefault}{\updefault}3(1-p)}}}}}
\put(1800,1854){\makebox(0,0)[lb]{\smash{{{\SetFigFont{12}{14.4}{\rmdefault}{\mddefault}{\updefault}3(1-p)}}}}}
\put(2775,5079){\makebox(0,0)[lb]{\smash{{{\SetFigFont{12}{14.4}{\rmdefault}{\mddefault}{\updefault}000}}}}}
\put(1275,3654){\ellipse{1050}{600}}
\put(2925,3654){\ellipse{1050}{600}}
\put(4725,3654){\ellipse{1050}{600}}
\put(2925,654){\ellipse{1050}{600}}
\put(1275,2154){\ellipse{1050}{600}}
\put(2925,2154){\ellipse{1050}{600}}
\put(4725,2154){\ellipse{1050}{600}}
\path(2850,4854)(2775,3954)
\path(2755.069,4076.077)(2775.000,3954.000)(2814.862,4071.094)
\path(1500,3954)(1502,3956)(1507,3959)
	(1515,3966)(1528,3976)(1546,3989)
	(1568,4006)(1594,4027)(1623,4050)
	(1655,4075)(1688,4101)(1722,4128)
	(1755,4154)(1787,4181)(1818,4206)
	(1848,4230)(1875,4253)(1901,4276)
	(1926,4297)(1948,4317)(1970,4336)
	(1990,4355)(2009,4374)(2027,4392)
	(2045,4411)(2063,4429)(2080,4448)
	(2096,4467)(2113,4486)(2130,4506)
	(2147,4527)(2164,4549)(2182,4573)
	(2201,4598)(2220,4624)(2240,4653)
	(2261,4683)(2283,4714)(2305,4747)
	(2328,4781)(2351,4814)(2373,4848)
	(2394,4880)(2414,4910)(2431,4936)
	(2445,4958)(2457,4976)(2475,5004)
\path(2435.344,4886.836)(2475.000,5004.000)(2384.874,4919.281)
\path(3075,3954)(3000,4854)
\path(3039.862,4736.906)(3000.000,4854.000)(2980.069,4731.923)
\path(3450,5079)(3452,5079)(3456,5078)
	(3463,5076)(3475,5074)(3491,5071)
	(3512,5067)(3539,5061)(3570,5055)
	(3607,5047)(3647,5038)(3692,5029)
	(3740,5018)(3790,5007)(3841,4995)
	(3893,4983)(3945,4970)(3996,4957)
	(4046,4944)(4093,4931)(4139,4917)
	(4181,4904)(4221,4890)(4259,4876)
	(4293,4861)(4324,4846)(4353,4831)
	(4379,4814)(4403,4797)(4425,4779)
	(4447,4758)(4467,4736)(4486,4711)
	(4504,4684)(4521,4654)(4537,4622)
	(4553,4587)(4568,4549)(4582,4509)
	(4596,4467)(4610,4422)(4624,4376)
	(4636,4330)(4649,4283)(4660,4236)
	(4671,4192)(4682,4149)(4691,4110)
	(4699,4074)(4706,4043)(4712,4017)
	(4716,3995)(4720,3979)(4725,3954)
\path(4672.049,4065.786)(4725.000,3954.000)(4730.883,4077.553)
\path(4350,3879)(4348,3881)(4343,3884)
	(4335,3891)(4322,3901)(4304,3914)
	(4282,3931)(4256,3952)(4227,3975)
	(4195,4000)(4162,4026)(4128,4053)
	(4095,4079)(4063,4106)(4032,4131)
	(4002,4155)(3975,4178)(3949,4201)
	(3924,4222)(3902,4242)(3880,4261)
	(3860,4280)(3841,4299)(3823,4317)
	(3805,4336)(3788,4354)(3770,4373)
	(3754,4392)(3737,4411)(3720,4431)
	(3703,4452)(3686,4474)(3668,4498)
	(3649,4523)(3630,4549)(3610,4578)
	(3589,4608)(3567,4639)(3545,4672)
	(3522,4706)(3499,4739)(3477,4773)
	(3456,4805)(3436,4835)(3419,4861)
	(3405,4883)(3393,4901)(3375,4929)
\path(3465.126,4844.281)(3375.000,4929.000)(3414.656,4811.836)
\path(1437,2455)(1362,3355)
\path(1401.862,3237.906)(1362.000,3355.000)(1342.069,3232.923)
\path(3450,729)(3452,729)(3456,730)
	(3463,732)(3475,734)(3491,737)
	(3512,741)(3539,747)(3570,753)
	(3607,761)(3647,770)(3692,779)
	(3740,790)(3790,801)(3841,813)
	(3893,825)(3945,838)(3996,851)
	(4046,864)(4093,877)(4139,891)
	(4181,904)(4221,918)(4259,932)
	(4293,947)(4324,962)(4353,977)
	(4379,994)(4403,1011)(4425,1029)
	(4447,1050)(4467,1072)(4486,1097)
	(4504,1124)(4521,1154)(4537,1186)
	(4553,1221)(4568,1259)(4582,1299)
	(4596,1341)(4610,1386)(4624,1432)
	(4636,1478)(4649,1525)(4660,1572)
	(4671,1616)(4682,1659)(4691,1698)
	(4699,1734)(4706,1765)(4712,1791)
	(4716,1813)(4720,1829)(4725,1854)
\path(4730.883,1730.447)(4725.000,1854.000)(4672.049,1742.214)
\path(4382,1878)(4380,1876)(4375,1873)
	(4367,1866)(4354,1856)(4336,1843)
	(4314,1826)(4288,1805)(4259,1782)
	(4227,1757)(4194,1731)(4160,1704)
	(4127,1678)(4095,1651)(4064,1626)
	(4034,1602)(4007,1579)(3981,1556)
	(3956,1535)(3934,1515)(3912,1496)
	(3892,1477)(3873,1458)(3855,1440)
	(3837,1421)(3820,1403)(3802,1384)
	(3786,1365)(3769,1346)(3752,1326)
	(3735,1305)(3718,1283)(3700,1259)
	(3681,1234)(3662,1208)(3642,1179)
	(3621,1149)(3599,1118)(3577,1085)
	(3554,1051)(3531,1018)(3509,984)
	(3488,952)(3468,922)(3451,896)
	(3437,874)(3425,856)(3407,828)
\path(3446.656,945.164)(3407.000,828.000)(3497.126,912.719)
\path(2400,5079)(2398,5079)(2394,5078)
	(2387,5076)(2375,5074)(2359,5071)
	(2338,5067)(2311,5061)(2280,5055)
	(2243,5047)(2203,5038)(2158,5029)
	(2110,5018)(2060,5007)(2009,4995)
	(1957,4983)(1905,4970)(1854,4957)
	(1804,4944)(1757,4931)(1711,4917)
	(1669,4904)(1629,4890)(1591,4876)
	(1557,4861)(1526,4846)(1497,4831)
	(1471,4814)(1447,4797)(1425,4779)
	(1403,4758)(1383,4736)(1364,4711)
	(1346,4684)(1329,4654)(1313,4622)
	(1297,4587)(1282,4549)(1268,4509)
	(1254,4467)(1240,4422)(1226,4376)
	(1214,4330)(1201,4283)(1190,4236)
	(1179,4192)(1168,4149)(1159,4110)
	(1151,4074)(1144,4043)(1138,4017)
	(1134,3995)(1130,3979)(1125,3954)
\path(1119.117,4077.553)(1125.000,3954.000)(1177.951,4065.786)
\path(2819,949)(2744,1849)
\path(2783.862,1731.906)(2744.000,1849.000)(2724.069,1726.923)
\path(3187,1849)(3112,949)
\path(3092.069,1071.077)(3112.000,949.000)(3151.862,1066.094)
\path(2398,724)(2396,724)(2392,725)
	(2385,727)(2373,729)(2357,732)
	(2336,736)(2309,742)(2278,748)
	(2241,756)(2201,765)(2156,774)
	(2108,785)(2058,796)(2007,808)
	(1955,820)(1903,833)(1852,846)
	(1802,859)(1755,872)(1709,886)
	(1667,899)(1627,913)(1589,927)
	(1555,942)(1524,957)(1495,972)
	(1469,989)(1445,1006)(1423,1024)
	(1401,1045)(1381,1067)(1362,1092)
	(1344,1119)(1327,1149)(1311,1181)
	(1295,1216)(1280,1254)(1266,1294)
	(1252,1336)(1238,1381)(1224,1427)
	(1212,1473)(1199,1520)(1188,1567)
	(1177,1611)(1166,1654)(1157,1693)
	(1149,1729)(1142,1760)(1136,1786)
	(1132,1808)(1128,1824)(1123,1849)
\path(1175.951,1737.214)(1123.000,1849.000)(1117.117,1725.447)
\path(2775,354)(2773,351)(2769,346)
	(2762,337)(2752,322)(2738,304)
	(2722,281)(2705,256)(2686,229)
	(2668,202)(2652,175)(2637,150)
	(2626,127)(2617,107)(2613,89)
	(2613,75)(2617,63)(2625,54)
	(2635,48)(2648,43)(2664,38)
	(2684,35)(2706,32)(2732,30)
	(2760,29)(2790,27)(2821,27)
	(2854,26)(2888,26)(2921,26)
	(2954,27)(2985,27)(3015,29)
	(3043,30)(3069,32)(3091,35)
	(3111,38)(3127,43)(3140,48)
	(3150,54)(3158,63)(3162,75)
	(3162,89)(3158,107)(3149,127)
	(3138,150)(3123,175)(3107,202)
	(3089,229)(3070,256)(3053,281)
	(3037,304)(3023,322)(3000,354)
\path(3094.397,274.067)(3000.000,354.000)(3045.676,239.049)
\path(2775,5454)(2773,5457)(2769,5462)
	(2762,5471)(2752,5486)(2738,5504)
	(2722,5527)(2705,5552)(2686,5579)
	(2668,5606)(2652,5633)(2637,5658)
	(2626,5681)(2617,5701)(2613,5719)
	(2613,5733)(2617,5745)(2625,5754)
	(2635,5760)(2648,5765)(2664,5770)
	(2684,5773)(2706,5776)(2732,5778)
	(2760,5779)(2790,5781)(2821,5781)
	(2854,5782)(2888,5782)(2921,5782)
	(2954,5781)(2985,5781)(3015,5779)
	(3043,5778)(3069,5776)(3091,5773)
	(3111,5770)(3127,5765)(3140,5760)
	(3150,5754)(3158,5745)(3162,5733)
	(3162,5719)(3158,5701)(3149,5681)
	(3138,5658)(3123,5633)(3107,5606)
	(3089,5579)(3070,5552)(3053,5527)
	(3037,5504)(3023,5486)(3000,5454)
\path(3045.676,5568.951)(3000.000,5454.000)(3094.397,5533.933)
\path(822,3565)(819,3564)(812,3561)
	(800,3557)(782,3551)(759,3543)
	(731,3533)(699,3522)(666,3511)
	(631,3501)(598,3491)(567,3483)
	(538,3477)(513,3474)(491,3473)
	(473,3476)(459,3481)(447,3490)
	(439,3501)(432,3514)(426,3531)
	(422,3551)(418,3574)(416,3600)
	(414,3627)(413,3657)(412,3687)
	(412,3719)(412,3750)(413,3780)
	(414,3809)(416,3837)(418,3862)
	(422,3884)(426,3903)(432,3919)
	(439,3931)(447,3940)(457,3946)
	(470,3949)(486,3949)(504,3945)
	(525,3939)(549,3930)(576,3919)
	(605,3905)(635,3891)(666,3875)
	(696,3859)(725,3844)(751,3829)
	(774,3817)(792,3807)(822,3790)
\path(702.807,3823.061)(822.000,3790.000)(732.388,3875.262)
\path(5250,3579)(5253,3578)(5260,3575)
	(5272,3571)(5290,3565)(5313,3557)
	(5341,3547)(5373,3536)(5406,3525)
	(5441,3515)(5474,3505)(5505,3497)
	(5534,3491)(5559,3488)(5581,3487)
	(5599,3490)(5613,3495)(5625,3504)
	(5633,3515)(5640,3528)(5646,3545)
	(5650,3565)(5654,3588)(5656,3614)
	(5658,3641)(5659,3671)(5660,3701)
	(5660,3733)(5660,3764)(5659,3794)
	(5658,3823)(5656,3851)(5654,3876)
	(5650,3898)(5646,3917)(5640,3933)
	(5633,3945)(5625,3954)(5615,3960)
	(5602,3963)(5586,3963)(5568,3959)
	(5547,3953)(5523,3944)(5496,3933)
	(5467,3919)(5437,3905)(5406,3889)
	(5376,3873)(5347,3858)(5321,3843)
	(5298,3831)(5280,3821)(5250,3804)
\path(5339.612,3889.262)(5250.000,3804.000)(5369.193,3837.061)
\path(750,2079)(747,2078)(740,2075)
	(728,2071)(710,2065)(687,2057)
	(659,2047)(627,2036)(594,2025)
	(559,2015)(526,2005)(495,1997)
	(466,1991)(441,1988)(419,1987)
	(401,1990)(387,1995)(375,2004)
	(367,2015)(360,2028)(354,2045)
	(350,2065)(346,2088)(344,2114)
	(342,2141)(341,2171)(340,2201)
	(340,2233)(340,2264)(341,2294)
	(342,2323)(344,2351)(346,2376)
	(350,2398)(354,2417)(360,2433)
	(367,2445)(375,2454)(385,2460)
	(398,2463)(414,2463)(432,2459)
	(453,2453)(477,2444)(504,2433)
	(533,2419)(563,2405)(594,2389)
	(624,2373)(653,2358)(679,2343)
	(702,2331)(720,2321)(750,2304)
\path(630.807,2337.061)(750.000,2304.000)(660.388,2389.262)
\path(5250,2079)(5253,2078)(5260,2075)
	(5272,2071)(5290,2065)(5313,2057)
	(5341,2047)(5373,2036)(5406,2025)
	(5441,2015)(5474,2005)(5505,1997)
	(5534,1991)(5559,1988)(5581,1987)
	(5599,1990)(5613,1995)(5625,2004)
	(5633,2015)(5640,2028)(5646,2045)
	(5650,2065)(5654,2088)(5656,2114)
	(5658,2141)(5659,2171)(5660,2201)
	(5660,2233)(5660,2264)(5659,2294)
	(5658,2323)(5656,2351)(5654,2376)
	(5650,2398)(5646,2417)(5640,2433)
	(5633,2445)(5625,2454)(5615,2460)
	(5602,2463)(5586,2463)(5568,2459)
	(5547,2453)(5523,2444)(5496,2433)
	(5467,2419)(5437,2405)(5406,2389)
	(5376,2373)(5347,2358)(5321,2343)
	(5298,2331)(5280,2321)(5250,2304)
\path(5339.612,2389.262)(5250.000,2304.000)(5369.193,2337.061)
\path(1168,3348)(1093,2448)
\path(1073.069,2570.077)(1093.000,2448.000)(1132.862,2565.094)
\path(4688,2448)(4613,3348)
\path(4652.862,3230.906)(4613.000,3348.000)(4593.069,3225.923)
\path(4981,3350)(4906,2450)
\path(4886.069,2572.077)(4906.000,2450.000)(4945.862,2567.094)
\path(1650,3429)(2700,2454)
\path(2591.651,2513.670)(2700.000,2454.000)(2632.478,2557.638)
\path(4500,3354)(3225,2454)
\path(3305.736,2547.711)(3225.000,2454.000)(3340.337,2498.693)
\path(2700,3354)(1725,2379)
\path(1788.640,2485.066)(1725.000,2379.000)(1831.066,2442.640)
\path(3150,3354)(4275,2379)
\path(4164.669,2434.921)(4275.000,2379.000)(4203.965,2480.262)
\path(1575,2454)(2550,3429)
\path(2486.360,3322.934)(2550.000,3429.000)(2443.934,3365.360)
\path(4425,2454)(3300,3429)
\path(3410.331,3373.079)(3300.000,3429.000)(3371.035,3327.738)
\path(2550,2379)(1575,3354)
\path(1681.066,3290.360)(1575.000,3354.000)(1638.640,3247.934)
\path(1541,1936)(1543,1934)(1548,1931)
	(1556,1924)(1569,1914)(1587,1901)
	(1609,1884)(1635,1863)(1664,1840)
	(1696,1815)(1729,1789)(1763,1762)
	(1796,1736)(1828,1709)(1859,1684)
	(1889,1660)(1916,1637)(1942,1614)
	(1967,1593)(1989,1573)(2011,1554)
	(2031,1535)(2050,1516)(2068,1498)
	(2086,1479)(2104,1461)(2121,1442)
	(2137,1423)(2154,1404)(2171,1384)
	(2188,1363)(2205,1341)(2223,1317)
	(2242,1292)(2261,1266)(2281,1237)
	(2302,1207)(2324,1176)(2346,1143)
	(2369,1109)(2392,1076)(2414,1042)
	(2435,1010)(2455,980)(2472,954)
	(2486,932)(2498,914)(2516,886)
\path(2425.874,970.719)(2516.000,886.000)(2476.344,1003.164)
\path(2475,3504)(2472,3503)(2465,3500)
	(2453,3496)(2435,3490)(2412,3482)
	(2384,3472)(2352,3461)(2319,3450)
	(2284,3440)(2251,3430)(2220,3422)
	(2191,3416)(2166,3413)(2144,3412)
	(2126,3415)(2112,3420)(2100,3429)
	(2092,3440)(2085,3453)(2079,3470)
	(2075,3490)(2072,3513)(2070,3539)
	(2069,3566)(2068,3596)(2068,3626)
	(2068,3658)(2069,3689)(2070,3719)
	(2072,3748)(2074,3776)(2076,3801)
	(2079,3823)(2083,3842)(2088,3858)
	(2093,3870)(2100,3879)(2109,3885)
	(2121,3888)(2135,3887)(2153,3882)
	(2173,3873)(2196,3862)(2221,3848)
	(2248,3831)(2275,3814)(2302,3796)
	(2327,3779)(2350,3764)(2368,3751)(2400,3729)
\path(2284.119,3772.262)(2400.000,3729.000)(2318.111,3821.705)
\path(3000,2454)(4350,3429)
\path(4270.283,3334.421)(4350.000,3429.000)(4235.154,3383.062)
\put(1125,3579){\makebox(0,0)[lb]{\smash{{{\SetFigFont{12}{14.4}{\rmdefault}{\mddefault}{\updefault}001}}}}}
\put(2775,3579){\makebox(0,0)[lb]{\smash{{{\SetFigFont{12}{14.4}{\rmdefault}{\mddefault}{\updefault}010}}}}}
\put(2775,2079){\makebox(0,0)[lb]{\smash{{{\SetFigFont{12}{14.4}{\rmdefault}{\mddefault}{\updefault}101}}}}}
\put(2775,579){\makebox(0,0)[lb]{\smash{{{\SetFigFont{12}{14.4}{\rmdefault}{\mddefault}{\updefault}111}}}}}
\put(1125,2079){\makebox(0,0)[lb]{\smash{{{\SetFigFont{12}{14.4}{\rmdefault}{\mddefault}{\updefault}011}}}}}
\put(4650,3579){\makebox(0,0)[lb]{\smash{{{\SetFigFont{12}{14.4}{\rmdefault}{\mddefault}{\updefault}100}}}}}
\put(4575,2079){\makebox(0,0)[lb]{\smash{{{\SetFigFont{12}{14.4}{\rmdefault}{\mddefault}{\updefault}110}}}}}
\put(4350,2754){\makebox(0,0)[lb]{\smash{{{\SetFigFont{12}{14.4}{\rmdefault}{\mddefault}{\updefault}1-p}}}}}
\put(5025,2754){\makebox(0,0)[lb]{\smash{{{\SetFigFont{12}{14.4}{\rmdefault}{\mddefault}{\updefault}p}}}}}
\put(1500,2754){\makebox(0,0)[lb]{\smash{{{\SetFigFont{12}{14.4}{\rmdefault}{\mddefault}{\updefault}p}}}}}
\put(750,2754){\makebox(0,0)[lb]{\smash{{{\SetFigFont{12}{14.4}{\rmdefault}{\mddefault}{\updefault}1-p}}}}}
\put(3525,2454){\makebox(0,0)[lb]{\smash{{{\SetFigFont{12}{14.4}{\rmdefault}{\mddefault}{\updefault}1-p}}}}}
\put(3600,3204){\makebox(0,0)[lb]{\smash{{{\SetFigFont{12}{14.4}{\rmdefault}{\mddefault}{\updefault}1-p}}}}}
\put(3225,3129){\makebox(0,0)[lb]{\smash{{{\SetFigFont{12}{14.4}{\rmdefault}{\mddefault}{\updefault}p}}}}}
\put(3525,1254){\makebox(0,0)[lb]{\smash{{{\SetFigFont{12}{14.4}{\rmdefault}{\mddefault}{\updefault}p}}}}}
\put(3225,1554){\makebox(0,0)[lb]{\smash{{{\SetFigFont{12}{14.4}{\rmdefault}{\mddefault}{\updefault}p}}}}}
\put(2475,4104){\makebox(0,0)[lb]{\smash{{{\SetFigFont{12}{14.4}{\rmdefault}{\mddefault}{\updefault}1-p}}}}}
\put(825,4254){\makebox(0,0)[lb]{\smash{{{\SetFigFont{12}{14.4}{\rmdefault}{\mddefault}{\updefault}1-p}}}}}
\put(4500,879){\makebox(0,0)[lb]{\smash{{{\SetFigFont{12}{14.4}{\rmdefault}{\mddefault}{\updefault}1-p}}}}}
\put(3375,54){\makebox(0,0)[lb]{\smash{{{\SetFigFont{12}{14.4}{\rmdefault}{\mddefault}{\updefault}1+p}}}}}
\put(3300,5604){\makebox(0,0)[lb]{\smash{{{\SetFigFont{12}{14.4}{\rmdefault}{\mddefault}{\updefault}1+p}}}}}
\put(4725,4254){\makebox(0,0)[lb]{\smash{{{\SetFigFont{12}{14.4}{\rmdefault}{\mddefault}{\updefault}p}}}}}
\put(1725,4254){\makebox(0,0)[lb]{\smash{{{\SetFigFont{12}{14.4}{\rmdefault}{\mddefault}{\updefault}p}}}}}
\put(3150,4104){\makebox(0,0)[lb]{\smash{{{\SetFigFont{12}{14.4}{\rmdefault}{\mddefault}{\updefault}p}}}}}
\put(3975,4254){\makebox(0,0)[lb]{\smash{{{\SetFigFont{12}{14.4}{\rmdefault}{\mddefault}{\updefault}1-p}}}}}
\put(2475,2754){\makebox(0,0)[lb]{\smash{{{\SetFigFont{12}{14.4}{\rmdefault}{\mddefault}{\updefault}1-p}}}}}
\put(3225,2754){\makebox(0,0)[lb]{\smash{{{\SetFigFont{12}{14.4}{\rmdefault}{\mddefault}{\updefault}p}}}}}
\put(5700,3804){\makebox(0,0)[lb]{\smash{{{\SetFigFont{12}{14.4}{\rmdefault}{\mddefault}{\updefault}1+p}}}}}
\put(75,3804){\makebox(0,0)[lb]{\smash{{{\SetFigFont{12}{14.4}{\rmdefault}{\mddefault}{\updefault}1+p}}}}}
\put(1875,1254){\makebox(0,0)[lb]{\smash{{{\SetFigFont{12}{14.4}{\rmdefault}{\mddefault}{\updefault}1-p}}}}}
\put(1200,954){\makebox(0,0)[lb]{\smash{{{\SetFigFont{12}{14.4}{\rmdefault}{\mddefault}{\updefault}p}}}}}
\put(2400,1554){\makebox(0,0)[lb]{\smash{{{\SetFigFont{12}{14.4}{\rmdefault}{\mddefault}{\updefault}1-p}}}}}
\put(0,2304){\makebox(0,0)[lb]{\smash{{{\SetFigFont{12}{14.4}{\rmdefault}{\mddefault}{\updefault}1+p}}}}}
\put(5700,2229){\makebox(0,0)[lb]{\smash{{{\SetFigFont{12}{14.4}{\rmdefault}{\mddefault}{\updefault}1+p}}}}}
\put(2925,5154){\ellipse{1050}{600}}
\path(2475,2004)(2472,2003)(2465,2000)
	(2453,1996)(2435,1990)(2412,1982)
	(2384,1972)(2352,1961)(2319,1950)
	(2284,1940)(2251,1930)(2220,1922)
	(2191,1916)(2166,1913)(2144,1912)
	(2126,1915)(2112,1920)(2100,1929)
	(2092,1940)(2085,1953)(2079,1970)
	(2075,1990)(2072,2013)(2070,2039)
	(2069,2066)(2068,2096)(2068,2126)
	(2068,2158)(2069,2189)(2070,2219)
	(2072,2248)(2074,2276)(2076,2301)
	(2079,2323)(2083,2342)(2088,2358)
	(2093,2370)(2100,2379)(2109,2385)
	(2121,2388)(2135,2387)(2153,2382)
	(2173,2373)(2196,2362)(2221,2348)
	(2248,2331)(2275,2314)(2302,2296)
	(2327,2279)(2350,2264)(2368,2251)(2400,2229)
\path(2284.119,2272.262)(2400.000,2229.000)(2318.111,2321.705)
\put(2625,3129){\makebox(0,0)[lb]{\smash{{{\SetFigFont{12}{14.4}{\rmdefault}{\mddefault}{\updefault}p}}}}}
\put(2025,3129){\makebox(0,0)[lb]{\smash{{{\SetFigFont{12}{14.4}{\rmdefault}{\mddefault}{\updefault}1-p}}}}}
\put(2175,2529){\makebox(0,0)[lb]{\smash{{{\SetFigFont{12}{14.4}{\rmdefault}{\mddefault}{\updefault}p}}}}}
\end{picture}
}
    \makeatother      
    \caption{Transition Probabilities (factorized by 3)}
\label{f:tp}
\end{center}
  \end{figure}
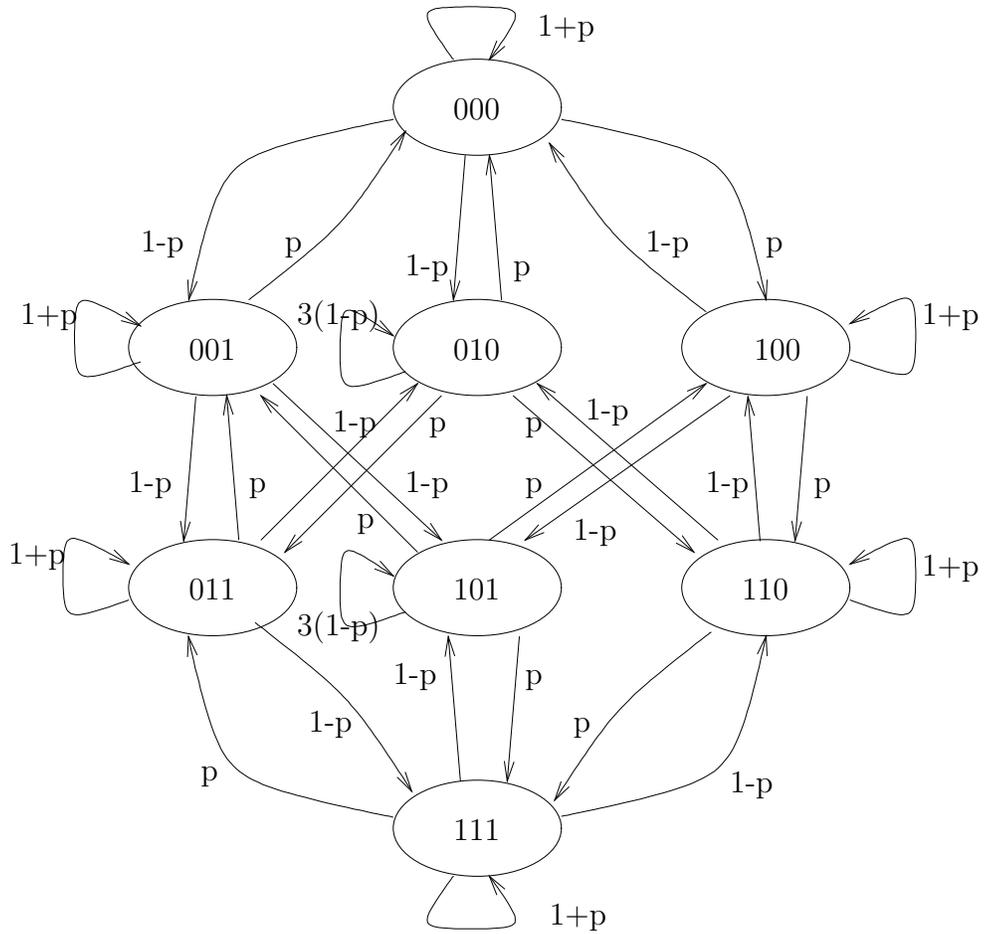

We continue describing our approach using 
a system of three processors and two states.
The transitions and probabilities of the system
appear in Figure \ref{f:tp}.

A read of a neighboring state returns with probability $p>1/2$ 
the correct value. Each configuration has four outgoing arrows,
one arrow for each state change of a processor, and one for
staying in the same state. There are two possibilities for
a state transition of a processor, one when the read
returns the correct value (probability $p$) and one when the read returns
a wrong value (probability $1-p$).
Since the daemon chooses to activate each processor with equal probability,
we have to use a factor $1/3$ for the above probabilities.

We now choose specific values for $p$ and compute 
powers of the probability matrix $\cal P$, such that 
the matrix in power $i$ and $i+1$ are equal
(${\cal P}^i={\cal P}^{i+1})$.
Then we conclude the percentage of being in a legal configuration
(not in the configurations 010 or 101). 
The following table shows different values for 
$p$ (1, 3/4, 1/2, 1/4) and the corresponding 
equilibrium vector ${\cal E}^T$; two figures
display the transition matrix ${\cal P}$ for 
the cases of $p=1$ and $p=3/4$.

\begin{center}
\begin{tabular}{|l|l|l|}
\hline
\textrm{Matrix} & p & \textrm{Equilibrium Vector} \\ \hline
\textrm{Fig}~\ref{f:tm1} & 1 & [1/6,1/6,0,1/6,1/6,0,1/6,1/6] \\ \hline
\textrm{Fig}~\ref{f:tm2} & 3/4 & 
[3/20,3/20,1/20,3/6,3/6,1/20,3/20,3/20] \\ \hline
 & 1/2 & [1/8,1/8,1/8,1/8,1/8,1/8,1/8,1/8] \\ \hline
 & 1/4 & [1/12,1/12,1/4,1/12,1/12,1/4,1/12,1/12] \\ \hline
\end{tabular}
\end{center}

The vectors show that the equilibrium probability for illegitimate
configurations is zero for the deterministic case, then 
increasing as $p$ reduces.  Clearly, we can investigate the behavior of
other systems with a range of probabilities,
using the same approach. The results can 
assist us in comparing different system designs.

\begin{lemma}  Dijkstra's algorithm is weakly stabilizing
when register reads are correct with probability $p>0$.
\end{lemma}

\begin{figure} [ht]
\begin{center}
\fbox{
\setlength{\textwidth}{6.00in}
\begin{minipage}{\textwidth}         

\begin{tabbing}
xxxxxxx \= xxxxxxx \= xxxxxxx \= xxxxxxx \= xxxxxxx \= xxxxxxx \= xxxxxxx \= \kill
2/3 \> 0 \> 0 \> 0 \> 1/3 \> 0 \> 0 \> 0 \\
1/3 \> 2/3 \> 0 \> 0 \> 0 \> 0 \> 0 \> 0 \\
1/3 \> 0 \> 0 \> 1/3 \> 0 \> 0 \> 1/3 \> 0 \\
0 \> 1/3 \> 0 \> 2/3 \> 0 \> 0 \> 0 \> 0 \\
0 \> 0 \> 0 \> 0 \> 2/3 \> 0 \> 1/3 \> 0 \\
0 \> 1/3 \> 0 \> 0 \> 1/3 \> 0 \> 0 \> 1/3 \\
0 \> 0 \> 0 \> 0 \> 0 \> 0 \> 2/3 \> 1/3 \\
0 \> 0 \> 0 \> 1/3 \> 0 \> 0 \> 0 \> 2/3 \\
\end{tabbing}

\end{minipage}
}
\caption[]{
\label{f:tm1}}
Transition Matrix for $P=1$
\end{center}
\end{figure}

\begin{figure} [ht]
\begin{center}
\fbox{
\setlength{\textwidth}{6.00in}
\begin{minipage}{\textwidth}         

\begin{tabbing}
xxxxxx \= xxxxxx \= xxxxxx \= xxxxxx \= xxxxxx \= xxxxxx \= xxxxxx \= \kill
7/4 \> 1/4 \> 1/4 \> 0 \> 3/4 \> 0 \> 0 \> 0 \\
3/4 \> 7/4 \>   0 \> 1/4 \> 0 \> 1/4 \> 0 \> 0 \\
3/4 \> 0 \> 3/4 \> 3/4 \> 0 \> 0 \> 3/4 \> 0 \\
0 \> 3/4 \> 1/4 \> 7/4 \> 0 \> 0 \> 0 \> 1/4 \\
1/4 \> 0 \> 0 \> 0 \> 7/4 \> 1/4 \> 3/4 \> 0 \\
0 \> 3/4 \> 0 \> 0 \> 3/4 \> 3/4 \> 0 \> 3/4 \\
0 \> 0 \> 1/4 \> 0 \> 1/4 \> 0 \> 7/4 \> 3/4 \\
0 \> 0 \> 0 \> 3/4 \> 0 \> 1/4 \> 1/4 \> 7/4 \\
\end{tabbing}

\end{minipage}
}
\caption[]{
\label{f:tm2}}
Transition Matrix for $p=3/4$ factorized by 3
\end{center}
\end{figure}

\newpage

\small{

}
\end{document}